# Soft Skills and Software Development: A Reflection from Software Industry

[1]Faheem Ahmed, [2]Luiz Fernando Capretz, [3]Salah Bouktif, [4]Piers Campbell

[1] *Department of Computing Science, Thompson Rivers University, Kamloops, BC, Canada*
[2] *Department of Electrical and Computer Engineering,*
*University of Western, London, Ontario, Canada*
[3,4] *College of Information Technology, UAE University, Al Ain, UAE*
fahmed@tru.ca, lcapretz@uwo.ca, salahb@uaeu.ac.ae, p.campbell@uaeu.ac.ae

## Abstract

*Psychological theories assert that not everybody is fit for every task, as people have different personality traits and abilities. Often, personality traits are expressed in people's soft skills. That is, the way people perceive, plan and execute any assigned task is influenced by their set of soft skills. Most of the studies carried out on the human factor in Information System concentrate primarily on personality types. Soft skills have been given comparatively little attention by researchers. We review the literature relating to soft skills and the software engineering and information systems domain before describing a study based on 650 job advertisements posted on well-known recruitment sites from a range of geographical locations including, North America, Europe, Asia and Australia. The study makes use of nine defined soft skills to assess the level of demand for each of these skills related to individual job roles within the software industry. This work reports some of the vital statistics from industry about the requirements of soft skills in various roles of software development phases. The work also highlights the variation in the types of skills required for each of the roles. We found that currently although the software industry is paying attention to soft skills up to some extent while hiring but there is a need to further acknowledge the role of these skills in software development. The objective of this paper is to analyze the software industry's soft skills requirements for various software development positions, such as system analyst, designer, programmer, and tester. We pose two research questions, namely, (1) What soft skills are appropriate to different software development lifecycle roles, and (2) Up to what extend does the software industry consider soft skills when hiring an employee. The study suggests that there is a further need of acknowledgment of the significance of soft skills from employers in software industry.*

## 1. Introduction

Software development has been considered a socio-technical endeavor for some time. Particularly in the case of software engineering, the need to communicate effectively with users and team members has been increasingly emphasized. Software is a by-product of human activities, such as problem solving capabilities, cognitive aspects, and social interaction. Human beings, however, are more complicated and less predictable than computers. Therefore the complexity of human personality gives rise to intricate dynamics in software development, ones that cannot be ignored but which have often been overlooked. When someone accepts the idea that awareness of psychological dimensions within oneself and human factors within one's work environment increase the software engineer's productivity, we must then wonder which psychological traits would be most worth investigating. No one can deny the fact that the production of any piece of software involves a human element, at least to some degree. Therefore it is equally intuitive that better results can be achieved if people with particular social skills are assigned to different phases of a project, those that best match their skills.

Software development has been roughly characterized as a set of activities comprising system analysis, design, coding, and testing. Logically they are different tasks which are coupled to achieve the objective of software construction and operation. The micro-level interpretation of these activities demands a set of abilities from the individuals involved in order to carry the activity out effectively. For example, the skills required to design a software system are quite dissimilar from those needed to test the software. The psychological hypothesis that not everyone can perform all tasks effectively suggests that personality traits play a critical role in the performance of people executing the same task. Although the field of software psychology was promulgated many decades ago [1], it has been neglected over the past years due to the complexity of human nature and how to address the issue of its impact on software skills and software development.

Over a period of time, the engineering of software has become an inter-disciplinary concept, which means that the skills necessary to successfully work in this area thirty years ago may no longer be applicable in today's workplace. For instance, software design has become much more than manipulating formal or semi-formal notations, but rather revolves around the interaction between designers and users: namely, the designer's perception of what the user wants, and the user's perception of what he/she really needs. Today, successful software is developed after a tremendous amount of time has been spent with the user in the form of prototyping, experimentation, and feedback. This is the *de facto* life cycle of many useful software systems. In contemporary society, the software industry has become a major employer. It has, in fact, generated many commentaries on the unique contributions of professionals engaged in its many sub-areas. Specialties within software engineering are as diverse as those in any other profession. Software engineering comprises stages in separate and distinct phases including, system analysis, design, programming, testing, and maintenance. It may be that certain personality types affect one phase but not others, or affect specific phases in different fashions.

The job requirements in the area of software engineering published online or offline normally divide the skill requirements into the two categories of "hard skills" and "soft skills". Hard skills are the technical requirements and knowledge a person should have to carry out a task. This includes the theoretical foundations and practical exposure an individual should have to successfully execute the planned task. Soft skills have their roots in psychology, and they cover a broad range of characteristics involving personality types, social interaction abilities, communication, and personal habits. In a working environment the people tend to weight that soft skills harmonize the hard skills. Soft skills are the personal individuality that has a major impact on the behavior of a person while having interaction with others in a work setup. In the collaborative environment of software development the possession and use of soft skills enhances the likelihood of an individual's success and contributes positively towards the common goal of the project. As companies rely more heavily on project teams and expect software engineers to partner with their customers, opportunities for conflicts abound. When it comes to soft skills, opposites do not always attract. When software professionals discuss how a task needs to be accomplished, they tend to be poor at verbalizing how the task affects the people involved. In fact, the greatest difference between software engineers and the general population is the percentage who take action based on what they think rather than what they feel, a fact that does not help in bringing software engineers closer to their customers.

Thus the time appears to be ripe to address specific psychological factors as applied to the work of various software industry professionals. The reason for addressing these human factors is largely the recognition that software engineers could benefit from greater awareness of themselves and others in order to develop their "soft skills", which in turn can positively influence their work. Software engineers have long realized that "soft skills" are increasingly required in the course of their work. However, specific studies have been sporadic and often incidental. Researchers from software engineering have explored the area of human factors in terms of personality traits or types only and no concrete work has been carried out in the area of those soft skills that actually help to demark a personality type indirectly. Studies to investigate what soft skills the software industry requires and what they ignore are missing completely. The objective of this study is to map the job requirements of different phases of software development life cycle with soft skills requirements sets, and provide up-to-date information about which soft skills are in high demand by the software industry for a number of job roles in the software development process and which soft skills are neglected by the software industry despite their importance. The main objective of this study is to investigate the answer to the following research questions:

**Research Question-I (RQ-I):** What soft skills are appropriate to different software development life cycle roles?

**Research Question-II (RQ-II):** Up to what extent does the software industry consider soft skills when hiring an employee?

## 2. Literature Survey

Developing high quality and efficient software has always been a high priority in the software industry. However, recruiting and retaining of high quality software engineering professionals has never been easy for management. Human factors in software engineering have different dimensions and studies have been performed from different perspectives. These perspectives could be the investigation of human factors in different phases of software life cycle, or the effect of teamwork in software development, or how a particular personality profile can suit a particular task like code review, or how it can concern other miscellaneous issues. Few studies have investigated the relationship between human skills and the software life cycle phases.

### 2.1 Survey of Human Factor in Software Engineering

Although it is important to understand the characteristics and personality traits of people involved in software development, little attention has been paid to these aspects [2]. Karn and Cowling [3] investigated the effects of different personality types on the working of a software engineering team. The study describes how ethnographic methods could be used to study software teams, and to understand the role of human factors in a software project. The results of the study indicated that certain personality types were more inclined to certain roles. Using the 16PF test [4], Acuna et al. [5] measured the correspondence between individual capabilities, such as intrapersonal, organizational, interpersonal, and management, and those of software roles, including team leader, quality manager, requirements engineer, designer, programmer, maintainer, tester, and configuration manager. Feldt et al. [6] evaluated the personality of 47 software professionals using the IPIP 50-item five-factor personality test [7]. After extensive statistical analyses, they found that there are multiple and significant correlations between personality factors and software engineering, and they concluded that individual differences in personality can explain and predict how judgments are made and how decisions are evaluated in software development projects. Hannay *et al.* [8] reported the impact of the Big Five [9] personality traits on the performance of pair programmers together with the impact of expertise and task complexity. The study involved 196 software professionals in three countries, forming 98 pairs. The results show that personality may be a valid predictor for long-term team performance; however, they found that personality traits have, in general, a modest predictive value for pair programming performance as compared with expertise, task complexity, and country. They recommend that more effort be spent on investigating other performance-related predictors such as expertise and task complexity, as well as other promising predictors, such as programming and learning skills. They also suggested that effort be spent on elaborating the effects of personality on various measures of collaboration, which in turn may be used to predict and influence performance. Insights into malleable factors such as learning, motivation, and programming may be used to improve pair programming performance. Within the field of software engineering, there are tremendous differences among individual performances in programming. Instructors of programming courses witness firsthand the huge variety among students in learning achievement and programming assignments. Furthermore, Shneiderman [10] reports that some programmers perform as much as ten times better than other programmers with similar backgrounds. Walz and Wynekoop [11] derive a methodology for identifying the traits and characteristics of top-performing software developers: (1) those that are best at making things work; (2) those who can best communicate with end-users, identify requirements, and transform them into a logical design; and (3) those destined for management. Turley and Bieman [12] also seek to identify the attributes that differentiate exceptional and non-exceptional software engineers and map them to the MBTI scale; these differences may result from the fact that productive people are carrying out the tasks they prefer. Cognitive styles have been examined as factors that may help to explain some of the variability; however, they have failed to consistently explain individual preference towards computer programming as opposed to another task, such as system analysis or design.

Bishop-Clark [13] investigated the relationship between cognitive aspects, personality traits and computer programming. She divided programming into several stages: problem representation, program design, implementation, and debugging. Moreover, she organized the theories and empirical studies of computer programming into four sub-tasks: problem solving, designing, coding, and debugging. The cognitive styles discussed entail dichotomies such as field dependency/independency, analytic/holistic, impulsivity/reflectivity, and divergent thinking; the personality traits include locus of control, and introversion/extroversion. These variables were mentioned because, according to her theory, they were all important within the realm of computer programming. In general, her model suggested that different characteristics are necessary for different tasks; for example, she indicated that the attribute necessary for debugging is reflectivity rather than impulsivity.

There is clear evidence that personality preferences have great impact on motivation, quality of work, and retention in the field of software engineering [14]. Hardiman [15] observed that the majority of good programmers were ISTJ, INTJ, ESTJ, ENTJ, ISFJ, or ENTP; in brief, they are mostly NTs and SJs. He also implies that NF types tend to have trouble with the sequential and abstract thinking necessary for writing programs. Capretz [16] has investigated the profile of a group of 100 software engineers (80% male and 20% female) who study in private or public universities, work for the government, or are employed by software companies. All these individuals are productive and motivated software engineers and were selected to participate in this study based on their occupation. This study has shown that ISTJ, ISTP, ESTP and ESTJ orientations compose over 50% of the sample and are therefore significantly over-represented, whereas the INFJ, ESFP and ENFJ groups are all particularly under-represented. Teague [17] tried to map the MBTI dimensions into three major subtasks of computing: system analysis, system design, and programming. A study of 38 computing professionals confirmed that computer specialists are not a homogeneous group. The personality types preferred for system analysts included those with a combination of extroversion (E) and intuition (N), as nine of the thirteen participants preferring analysis had these two characteristics. Furthermore, ten of the thirteen participants favoring system design fell within the most preferred range of attributes, with six having the intuition (N) and thinking (T) characteristics considered useful for higher level design, two having the sensing (S) and Judging (J) characteristics desired for dealing with the detailed aspects of design, and two being ISTP – practical problem solvers. Finally, seven out of the ten respondents preferring programming also corresponded to personality types of traditional programmers, such as ISTJ, who are sensors (S) with attention to detail; the other three were classified as intuitives (N) and thinking (T).

More recently, Capretz and Ahmed [18] have mapped some opposing psychological traits, such as extroversion-introversion, sensing-intuition, thinking-feeling, and judging-perceiving, to the main stages of a software development life cycle. Subsequently, they have argued that assigning a person with specific psychological characteristics to the stage of the software life cycle best suited for his or her traits increases the chances of a successful outcome for the project. Software is developed by people, used by people, and supports people's work. As such, human characteristics, behavior, and cooperation are central to practical software development [19]. Specifically, the personality of software engineers has become increasingly important in recent years. First, today's software engineers are expected to have a broader range of skills than in the past. Secondly, many users are dissatisfied with the personal rather than the technical services they receive from software engineers. Several studies involving the personality traits of software engineers have been reported. However, much of the research has sought to classify software engineers into personality profiles according to particular psychological attributes measured by a personality instrument. A common thread running through the results of these and other similar studies is the prevalence of introverts (I), thinking (T), judging (J), and almost as many sensing (S) as intuitive (N) types among software professionals. While these empirical studies suggest that the MBTI polls are related to software engineering, they do not specify at which phase of the software life cycle they occur or how they are related.

### 2.2  Survey of Soft Skills and Software Development

The advertisement of jobs in the area of software development generally divides the skill requirements into two categories of "hard skills" and "soft skills". The need for research on skill sets of software developers in general and software designers in particular is not well supported by either academic or industrial researchers. Human factors have long been recognized as a critical factor in software development, and much of the research has been carried out with regard to personality traits. Technical skills are defined as "those skills acquired through training and education or learned on the job and are specific to each work setting," while soft skills are defined as "the cluster of personality traits, social graces, language skills, friendliness, and optimism that mark each one of us to varying degrees" [20]. Hard skills are the technical requirements and knowledge a person should have to carry out a task. This includes the theoretical foundations and practical exposure an individual should have to successfully execute the planned task. Even though soft skills are the psychological phenomena which cover the personality types, social interaction abilities, communication, and personal habits, people believe that soft skills complement the hard skills. Soft skills refer to the cluster of personality traits and attitudes that drives one's behavior [21]. They indirectly define the personality traits, ability to have social interaction, and eagerness which individuals acquire as they grow and mature. Other classical sets of soft skills are: active listening, negotiating, conflict resolution, problem solving, reflection, critical thinking, ethics, and leadership skills [22] [23] [24] [25]. Soft skills complement the technical skills requirements of a job [26]. According to Goleman [27] the possession and use of soft skills contributes more to an individual's ultimate success or failure than technical skills or intelligence. Bolton [28] reports that 80 percent of individuals who fail at work do not fail due to their lack of technical skills but rather because of their inability to relate well with others. A lack of soft skills is more likely to get an individual's employment terminated than a lack of cognitive or technological skills [29]. As software development is a team-based endeavor, it is necessary to determine which soft skills are required in which phase of software development. These skills vary, depending upon the needs and characteristics of the software project and domain of application, but there are some skills that may be common to a specific phase of software development. McGee [30] found, for example, that 68% of CIOs said that "soft skills," i.e., skills of a non-technical nature, such as communication and team building, are more important today than five years ago. Young and Lee [31], and Van Slyke [32] found that employers tend to rate non-technical skills higher than technical skills. Cappel [33] concluded that non-technical skills such as oral and written communications, problem solving, and the ability to learn apply to virtually all IS jobs. According to the survey of Green [34] the job of systems analyst considers behavioral skills such as diplomacy, politics, and sales, more important, while users consider technical skills such as programming more important. Khan and Kukalis [35] concluded that both hard and soft skills are important, but the hard skills are considered less important than soft skills. Leitheiser [36] found that people-oriented and organizational skills were more important than technical skills. Trauth *et al.* [37] examined the perceived importance of skills for information systems professional and their academic preparation. Leitheiser's [38] survey of information systems managers ranked interpersonal communication skills as most important. Wade and Parent [39] found that analysts perceive organizational skill as most important, whereas Green [40] found that behavioral skills such as diplomacy, sales and politics, are most important. Litecky *et al*. [41] presented an overview of studies dealing with the paradox of soft skills versus technical skills in hiring. Conversely, some surveys such as [40] [42] [43] concluded that technical skills are most important in affecting IS success.

## 3. Theoretical Foundations of the Study

Software is a term with a somewhat fuzzy connotation. People who talk about software may be thinking about the structure of a program, the functionality of an application system, the look and feel of an interface, or the overall user experience with a hardware-software environment. Software engineering spans both new software developments and the maintenance of legacy systems, and each software life cycle phase defines its own contexts of understanding about what matters, what can be designed, and what tools and methods are appropriate. Indeed, effective software development demands a broad set of skills. If a project lacks people with certain preferences, some individuals may have to perform tasks to which they are not naturally suited or which they may not find enjoyable. Despite early interest in the importance of human factors in the engineering of software, particularly the personal characteristics of people involved in the software engineering processes, such factors have been largely overlooked. This oversight, in turn, has prevented the acquisition of detailed knowledge about how different aspects of personality correlate with software life cycle phases and has hindered the use of this knowledge to improve software engineering processes. A more focused approach may help identify which software life cycle phase a particular personality type has the most significant impact upon. Ang and Slaughter [44] observe that despite the increasing importance of soft skills, very little systematic research has investigated such skills and, to an even less degree has measured these skills.

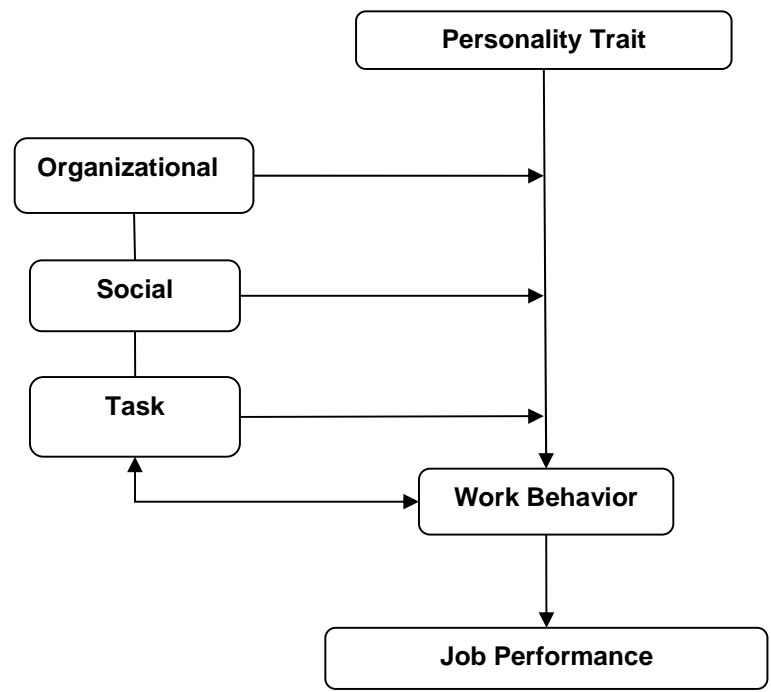

**Figure 1.** A Personality Trait-based Model of Job Performance [49]

Wagber and Sternberg [45] observe that graduates who performed well academically in schools did not necessarily perform well in the workplace. According to Joseph *et al.* [46] there is a growing awareness that technical skills alone are insufficient for success in IT. Job performance is linked with personality traits because personality traits further define the work attitude of an individual. Campbell [47] presents a performance prediction model proposing that performance is determined directly by a combination of knowledge, skill, and motivation. Motowidlo *et al.* [48] Suggested that personality variables contribute to performance by way of habits, skills, and knowledge. A personality trait-based job performance model proposed by Tett and Burnett [49], is only partially redrawn (for simplicity reason) in Figure 1. Fishbein and Ajzen's [50] theory of reasoned action postulates that intentions to perform a behavior are a function of an individual's attitude toward the behavior and subjective norms about what relevant others think about the behavior. According to Figure 1, job performance is influenced by the work behavior of an individual, and it ultimately reflects the personality characteristics of the employee. In Figure 1, "task" defines the activities to be performed in the job; for example in the case of requirements engineer one has to interact with customers to capture product requirements. In this case, if the employee has good communication skills along with technical expertise, the likelihood of success in job is higher. "Social" defines the characteristics of an individual to interact with others, for example a characteristic such as interpersonal skills. Most of software development involves teamwork in which conflict is a common phenomenon. An employee with better interpersonal skills has a tendency to resolve conflicts. In Figure 1, "organization" illustrates the organizational culture. Organizational culture has been characterized by many authors as comprising a set of shared values, beliefs, assumptions, and practices that shape and guide the attitudes and behaviors of individuals within the organization [51][52][53]. Lee *et al.* [54] found that the performance of an individual is influenced by their personality and abilities. They found that information system development requires developer skills and abilities that may go beyond traditional technical expertise. They further highlighted that the interplay of emotional intelligence (a terminology currently used for soft skills), creativity and job performance plays a significant role in information system development. O' Reilly [55] observes that when different personalities interact with environment they produce different responses. Chilton *et al.* [56] found that the person job fit phenomenon is positively related to the job performance of software developers. According to Schutte *et al.* [57] emotional intelligence is related to the quality of interpersonal relationship, and having higher emotional intelligence leads to better job performance.

The theoretical foundations of the study shown in Figure 2 depict that, in general, job advertisements in industry define requirements in two broad categories: technical and soft skills. Technical skills are related to the tasks or activities to be performed in carrying out one's role and demonstrate the knowledge and expertise required. Soft skills, on the other hand, define the emotional and behavioral set of requirements. These soft skills are in turn linked to the social requirements of a given position - for example, the need to communicate effectively with internal team members or external stakeholders. Alternatively, a position may require more solitary individuals who will not be involved in a collaborative role and who are capable of working alone and under minimal supervision. These two sets of requirements (technical and soft skills) are influenced by the inherent personality traits of the individual, and this aspect of personality make-up can further characterize the behavior of a given individual in a work setting. Workplace behavior is one of the critical factors in the job performance of the employee, and effectively represents the balance between the technical and social (soft) aspects of a given role.

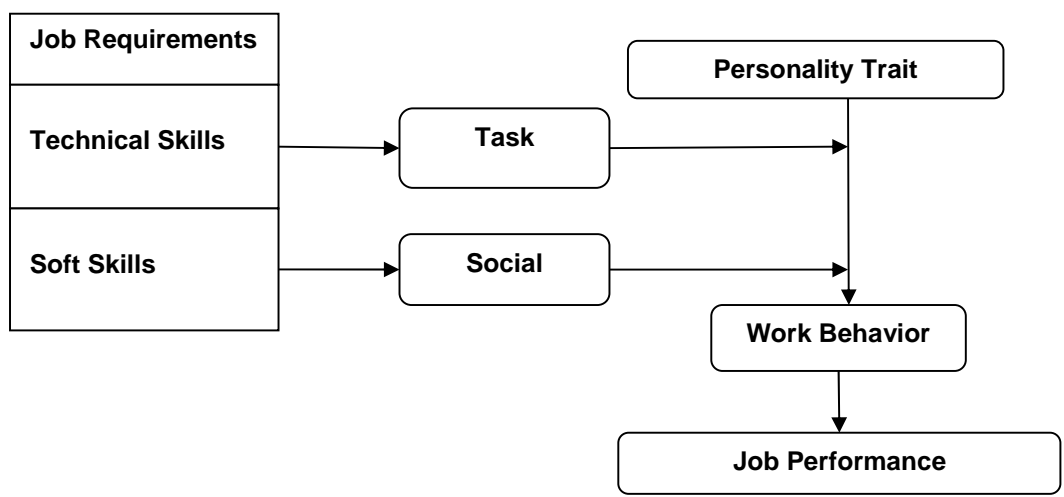

**Figure 2.** Research Model of the Study

### 3.1 Research Methodology

The methodology revolves around the foundations of empirical research in software engineering. The format of the research questions RQ-I and RQ-II are "descriptive and classification questions" and "existence question" respectively as specified in [58]. In order to find out answer to the research question RQ-I we used literature survey and expert opinion techniques. We did a comprehensive literature survey in the areas of software engineering and human factors, soft skill in software development and job performance modeling which provides us foundations to further investigate answer to RQ-I. We selected from on-line job portals most commonly used job requirements of individual roles in software development such as system analysts, designer, programmer and tester, and then map the possible soft skills required to carry out these tasks. We consulted with some experts in these areas to "read between the lines" of the job requirements and provide us their judgment about possible soft skill requirement to carry out that task. We compiled the independent opinion of the experts and used our own judgment and guidance from literature survey in making final mapping. In case of RQ-II we collected data from software industry about soft skills requirements in the job advertisements and compiled the data to interpret and find out answer to RQ-2.

## 4. Survey Setup and Data

Surveys, experiments, metrics, case studies, and field studies are all examples of empirical methods used to investigate both software engineering processes and ethical issues, which also must be taken into account [59]. In this survey of the software job market we analyzed the soft skills required by the software industry for various titles of software development jobs. The survey consists of looking at 650 jobs advertised across the globe. For this study we visited some of the leading online job portals such as workopolis.ca (North America), eurojobs.com (Europe), monsterindia.com (Asia), and seek.com.au (Australia). Figure 3 illustrates the geographical distribution of the subjects. The geographical distribution of the dataset covers 32% (North America), 23% (Europe), 25% (Asia) and 20% (Australia). We examined jobs advertised under the exact and related titles of "system analyst", "software designer", "computer programmer", and "software tester". The examples of related tittles for computer programmer are such as "Software Developer", "Coder", and "Application Developer". Similarly in case of system analyst some of the related titles were "Business Analyst", "Requirements Analyst", "Technical Analyst", and "Application Analyst". Figure 3 also illustrates the job classifications of the subjects. The total dataset of 650 jobs covers 24% (System Analysts), 25% (Software Designer), 26% (Computer Programmer) and 25% (Software Tester).

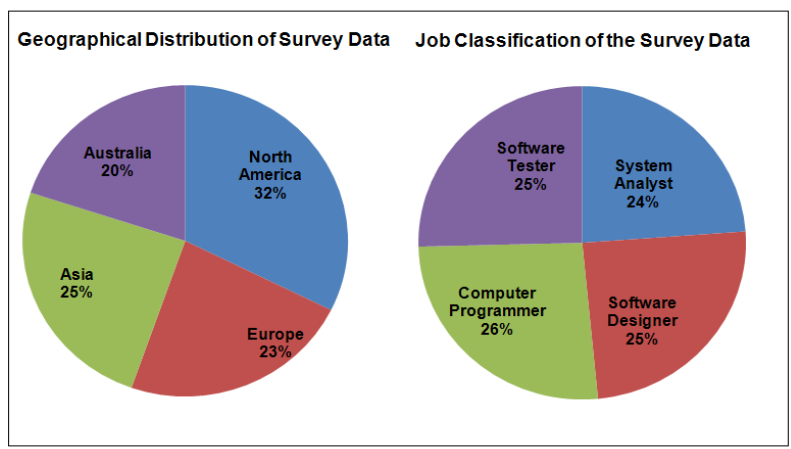

**Figure 3.** Geographical Distribution and Job Classification of the Survey Data

Figure 4 summarizes of the data collected for this survey with respect to geographical location (North America, Europe, Asia, and Australia) and job category (system analysts, software designer, computer programmer and software tester). We collected data concerning seven soft skills, which include communication skills, interpersonal skills, analytical and problem solving skills, team player, organizational skills, ability to work independently, and open and adaptable to changes. We found these seven soft skills most commonly used in the advertisement of the jobs. The requirement of soft skills in software industry for various roles is described using a three point scale of "High Demand (> 66%)", "Moderate Demand (> 33% and ≤ 66)", and "Low Demand (≤ 33%). It is also important to mention here that when we visited a job page the inclusion of the job post in the study dataset depends on the categorically presence of at least one of the above mentioned soft skills in the job posting. Therefore we looked specifically for these key words of the soft skills.

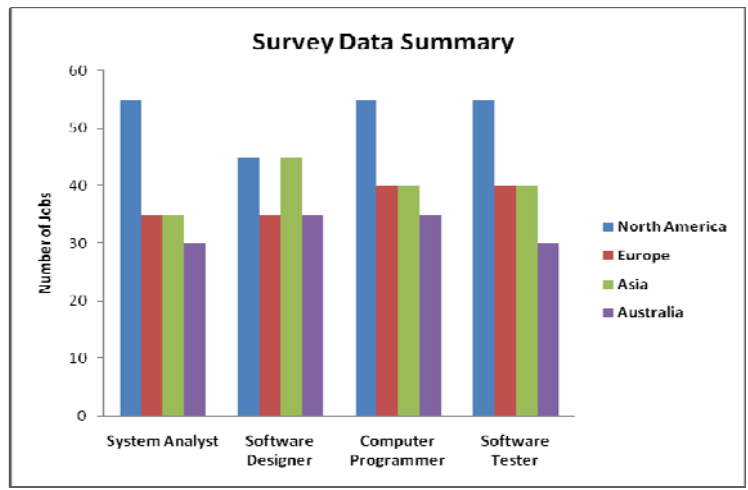

**Figure 4.** Data Summary of the Survey Geographical and Job Classification

In order to better understand the usage and significance of these soft skills we provide some elaboration of these concepts. Communication skills are the set of skills that enables an individual to convey information so that it is received and understood. The term "interpersonal skills" refers to the measure of a person's ability to deal with other people through social communication and interactions under both favorable and unfavorable conditions. Analytical and problem solving skill is the ability to visualize, articulate, and solve complex problems and concepts, and make decisions that make sense based on available information. Such skills include the demonstration of the ability to apply logical thinking to gathering and analyzing information, designing and testing solutions to problems, and formulating plans. Team player is a characteristic of an individual who can work effectively in a team environment and contributes to the desired goal. . Organizational skills refer to the ability of individual managing tasks within time, resource and sequence of execution. The ability to work independently denotes the individual's capability of carrying out a task with minimal supervision. Open and adaptable to changes reflects the ability of an individual to accept changes in the carrying out of tasks without showing resistance.

## 5. Survey Results and Discussion

The results of this investigation will be discussed separately, based on the specific role performed by a software engineer, that is, as a system analyst, designer, programmer, or tester. In this section each role is examined by analyzing the job requirements of the role and the possible mapping of the soft skills required. We collected the job requirements of the individual roles from the online job portals. Later we present the data collected about soft skills required by the software industry; we do so in order to understand the extent to which the software industry is paying attention to this phenomenon.

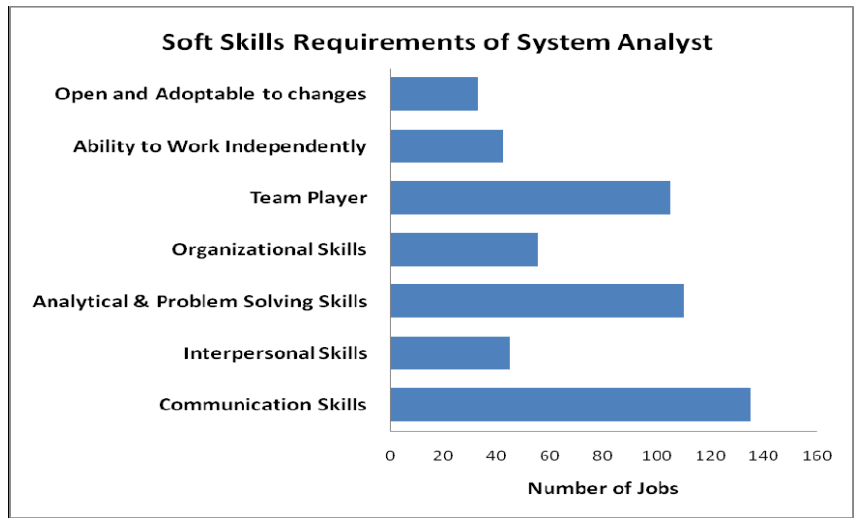

**Figure 5.** Soft Skills Qualificatons of System Analyst

**Table 1.** System Analysts' Job and Soft Skills Requirements

| Sample System Analysts Job Requirements | Possible Soft Skills Requirements |
|---|---|
| Liaising extensively with external or internal clients. | Communication Skill, Interpersonal Skills |
| Analyzing clients' existing systems. | Analytical & Problem solving Skills, Open and Adaptable to Changes |
| Translating client requirements into highly specified project briefs. | Communication Skills, Analytical & Problem Solving Skills, Open and Adaptable to Changes |
| Identifying options for potential solutions and assessing them for both technical and business suitability. | Analytical & Problem Solving Skills |
| Creating logical and innovative solutions to complex problems. | Analytical & Problem solving Skills, Open and Adaptable to Changes |
| Drawing up specific proposals for modified or replacement systems. | Communication Skills, Analytical & Problem Solving Skills, Open and Adaptable to Changes |
| Producing project feasibility reports. | Communication Skills, Analytical & Problem Solving Skills |
| Working closely with developers and a variety of end users to ensure technical compatibility and user satisfaction. | Interpersonal Skills, Communication Skills |
| Overseeing the implementation of a new system. | Organizational Skills, Interpersonal Skills |
| Planning and working flexibly to a deadline. | Organizational Skills, Open and Adaptable to Changes |

## 5.1 System Analyst

### 5.1.1. Theoretical Foundation

The broader job description of a system analyst is a role that requires the identification of high-level components in a real-world application and the decomposition of the software system [60]. Among other further micro-level subset of skills a system analyst should posse are determining the user/client's needs, considering the requirements expected to be satisfied by the software system, understanding the system's essential features, and creating an abstract model of the application by which these requirements are met.

This requires a high level of analytical and problem-solving skills. The main product of the system analysis phase is a graphical and/or textual description (informal or formal) of an abstract model of the application. System analysis requires a great deal of human interaction with users and clients; for such interactions, communication skills are essential. Additionally, system analysts must be able to empathize with the customer's problems to fully understand their needs; hence interpersonal, analytical, the ability to listen actively are highly desirable skills. These can be critical for software professionals, who are often viewed as being disconnected from users. There appears to be a tendency for software engineers to assume that because they have technical expertise and most users do not, their solutions are the most appropriate, even though the user may not immediately appreciate this fact. Table 1 illustrates the most common job requirements of system analyst advertised in the online portals. After mapping the soft skills required in carrying out these activities we found that communication, analysis, problem solving, and being open and adaptable to changes as well as having good interpersonal skills is all highly desirable for the job of system analyst.

### 5.1.2. Reflection from Software Industry

In the collection of 155 system analyst opportunities from North America, Europe, Asia and Australia, the analysis shows that high demand soft skills include communication skills (87%), analytical and problem solving (70%) and team player (67%). There is moderate demand for organizational skills (35%). The least required soft skills for a system analyst are interpersonal skills (29%), ability to work independently (27%), and being open and adaptable to changes (21%). Figure 5 summarizes the survey results. The analysis shows that although some of the vital soft skills, ones such as communication, analysis, problem solving, and operating as a team player are in high demand. Others, such as interpersonal skills and innovation are not given much attention by the software industry. The job of system analyst requires much of interaction with the user; therefore industry should pay attention to the interpersonal skills for this title of the software engineering job. The analysis activity is somehow supports the concept of iterative development where you understand the problem through inception elaboration and repetitive cycle. This iterative nature of the system analysis process requires a personality having characteristics of being open and adaptable to changes. The analysis of job requirements of system analyst shows that there is a moderate demand of being open and adaptable to change. Keeping in view the significance of this skill, the software industry should pay more attention to it because change management has been a key issue in software requirements management.

## 5.2 Software Designer

### 5.2.1. Theoretical Foundation

Although the term design is ambiguous, a "good" design leads to success while a "poor" design points to disaster, a fact which makes the design process crucial. Although there is significant diversity among design principles, we can find common traits that are applicable to the design of any artifact, whether it is a poster, a household appliance, or a housing development. Software design is still in the evolution phase and far from having a consensus on its relevant principles, but it requires the human creativity possessed by other disciplines such as architecture, marketing, or graphic design, rather than the hard-edged formulaic certainty of engineering design. Software design is an exploratory process: the designer looks for components by trying out a variety of schemes in order to discover the most natural and reasonable way to refine the solution [61]. This requires analytical and problem solving to understand and provide an appropriate solution.

There has been a tendency to present software design in such a manner that it looks easy to do. Nevertheless, in the design of large and complex software, identification of key components is likely to take some time. Repetitions are not unusual, as achieving a "good" design usually takes several iterations. The number of iterations also depends on the designer's insight and experience in the application domain. Software designers should have the ability to see the "big picture", the ability to single out the items that are relevant from imprecise and large quantities of fuzzy data, a task which requires an ability to identify patterns. Naturally, there is a tendency towards intuition, and normally those who are imaginative and innovative thrive during design compared with their fact-oriented, black-and-white minded counterparts.

Soft Skills and Software Development: A Reflection from Software Industry
Faheem Ahmed, Luiz Fernando Capretz, Salah Bouktif, Piers CampbellSoftware designers have a wide range of tasks, which include: prototyping, elaborating processing functions, and defining inputs and outputs, all of which require communication, interpersonal, analytical and problem solving and organizational skills. Table 2 shows some of the specific tasks required to carry out a software designer job in an organization. These activities are commonly listed in the job advertisements on the job portals. We have mapped the possible soft skills required to carry out the job. After mapping the soft skills required in carrying out these activities we found that communication, analytical and problem solving, organizational skills, open and adaptable to changes and interpersonal skills are highly desirable for the job.

**Table 2.** Software Designers' Job and Soft Skills Requirements

| Sample Software Designer Job Requirements | Possible Soft Skills Requirements |
| --- | --- |
| Ability to craft scenarios, storyboards, information architectures, features and interfaces. | Communication Skills, Analytical & Problem Solving Skills, Organizational Skills |
| Collaborating closely with management, engineers and fellow designers to evaluate and iterate on ideas and designs. | Communication Skill, Interpersonal Skills, Organizational Skills, Team Player |
| Prototyping user experience and design ideas. | Communication Skills, Analytical & Problem Solving Skills |
| Understanding business opportunities and assisting project team with respect to architecture of the technical solution. | Communication Skills , Problem Solving Skills |
| Creating an architectural design with the necessary specifications for the hardware, software, and data. | Communication Skills, Analytical & Problem Solving Skills, Organizational Skills |
| Working closely with system users to ensure that implementation meets customer requirements. | Communication Skill, Interpersonal Skills |
| Developing, documenting and revising system design procedures. | Communication Skills, Analytical & Problem Solving Skills, Organizational Skills |
| Participating in testing and evaluating the systems functionality to ensure successful integration. | Team Player, Interpersonal Skills |
| Determining hardware, software and network requirements of the software system. | Analytical & Problem Solving Skills |

### 5.2.2. Reflection from Software Industry

We collected the data of 160 software designer roles from the job market, and analysis shows that high demand soft skill is communication skills (88%). There is moderate demand for analytical and problem solving (37%) and team player (45%). The least required soft skills for software designers are interpersonal skills (25%), organizational skills (18%), ability to work independently (15%), open and adaptable to changes (6%). Figure 6 summarizes the results. Software design activity requires a great deal of organizational skills which make for working efficiently and effectively. Thinking about software design in its entirety can be overwhelming and complex. But by breaking the design down into smaller and more manageable pieces (i.e. organizing them) they do not seem to be as difficult to achieve. The organizational skills you apply toward planning each day insure that you are at least somewhat productive and that you accomplish what you must in order to avoid delaying the project. The analysis of job requirements and soft skills in the case of a software designer illustrate that organizational skill is one of the least required skill sets, which seems contrary to the nature of the designer job. Therefore the software industry should pay more attention to this vital skill set. Sometimes a design goes through many cycles of changes and prototyping in order to ensure that implementation meets customer requirements. Defining features and interfaces in the architecture requires evaluating and iterating ideas. The iterative nature of the software designing process requires a personality having the characteristics of openness and adaptability to changes, but this requirement has not been currently supported by the software industry and the skill sets of openness and adoptability to changes is also one of the least required skills set in demand. The software industry should pay more attention to this significantly important skill set.

182

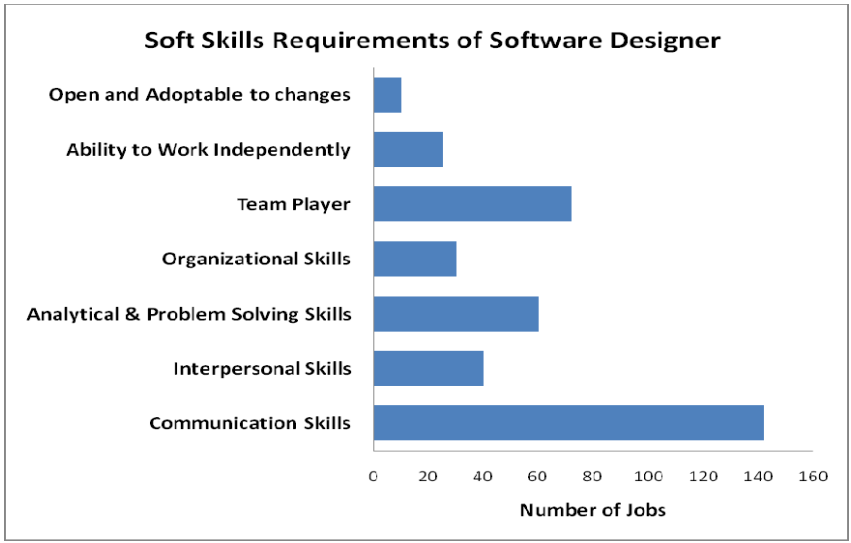

**Figure 6.** Soft Skills Qualifications of Software Designer

**Table 3.** Software Programmers' Job and Soft Skills Requirements

| Sample Software Programmer Job Requirements | Possible Soft Skills Requirements |
|---|---|
| Participates in development efforts, elaborates and documents all business-related applications. | Communication Skills, Team Player |
| Analyzes business requirements for system subcomponents and prepare detailed programming specifications for assigned system applications. | Communication Skills, Analytical & Problem Solving Skills, Organizational Skills, Open and adaptable to changes |
| Formulates test cases to test application software in development, to ensure a program's functionality matches its specification's business requirements. | Analytical & Problem Solving Skills, Organizational Skills, open and adaptable to changes |
| Analyzes technical specifications, builds and implements functionally accurate and modular application programs according to approved design specifications. | Analytical & Problem Solving Skills, Organizational Skills, Communication Skills |
| Coordinates programming tasks, team members, and projects within the department. | Communication Skills, Team Player, Interpersonal Skills |
| Determines forms, procedures, and other documentation needed for installation and maintenance of application programs. | Analytical & Problem Solving Skills, Organizational Skills |
| Translates detailed flow charts into coded machine instructions and conferring with technical personnel in planning programs. | Communication Skills, Analytical & Problem Solving Skills, Organizational Skills, Ability to Work Independently |
| Selects and incorporates available software programs. | Analytical & Problem Solving Skills, Organizational Skills |

## 5.3  Software Programmer

### 5.3.1.  Theoretical Foundation

The simplest definition of a programmer's job is translating a refined version of the design into a computer executable program [62]. The broader technical aspects of this phase requires the identification of control structures, relevant variables and data structures, and a detailed understanding of the syntax and specifics of a given programming language, which in turn requires analytical and problem solving skills. This usually follows an iterative stepwise refinement process that is mostly top-down, breadth first. Therefore programmers should attend to details and maintain an open, logical and analytical thinking process. The problem of interpreting and giving meaning to variables may require analytical thinking especially. Software programmers need to translate a design into executable code, a task which needs understanding of the design and a number of judgment decisions. The code structure is largely dependent upon how the programmer develops a mental representation of the design. The rapid development in programming paradigms requires a programmer to keep himself/herself up-to-date with the latest state of the art techniques. Programming tasks, such as determining details of module logic, establishing file layout, and coding programs, show little requirement for interpersonal contact and reveal the programmer's work life as essentially a solitary one, which requires organizational skill, fast-learner ability, and a capacity to adapt to changes. Programmers often work independently but share their code during inspection and module integration, a fact which requires the ability to work with others as well. In table 3 the software programmers' general job activities are related with possible soft skills requirements. After mapping the soft skills required in carrying out these activities we found that communication, analytical and problem solving, open and adaptable to changes and organizational skills are highly desirable for the job.

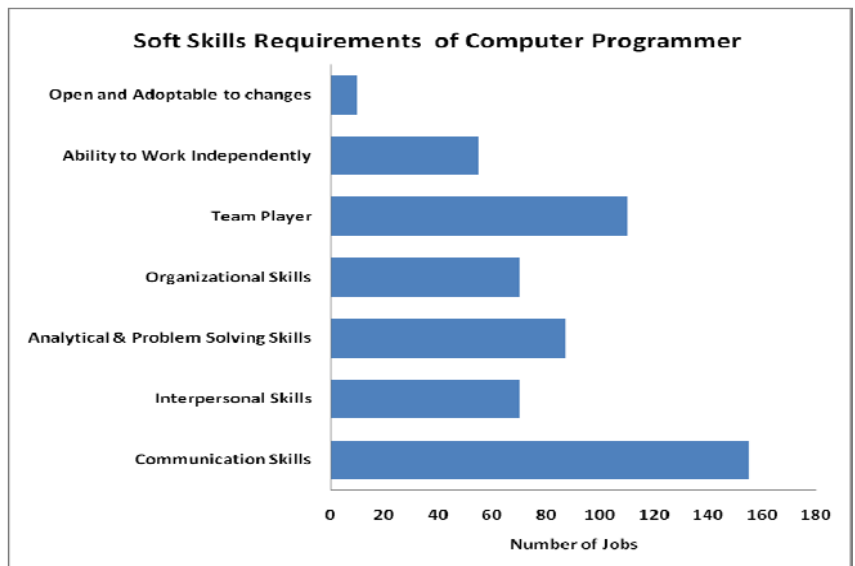

**Figure 7.** Soft Skills Desirable for a Computer Programmer

### 5.3.2. Reflection from Software Industry

We collected the data of 170 computer programmers from the software job market, and analysis shows that the soft skill in high demand is the ability to communicate (communication skills 91%). There is moderate demand for analytical and problem solving (51%), being a team player (64%), organizational skills (41%) and interpersonal skills (41%). The least required soft skill for computer programmers are ability to work independently (32%), open and adaptable to changes (5%). Figure 7 summarizes the survey results. Programming mainly requires problem solving and a lot of analytical skills. The structure of the program requires logical thinking at various locations and may increase the quality of the software; for example, where to introduce recursion, choice of loop, their termination criteria, dividing the program into smaller functions, etc. Currently there is a moderate demand of analytical and problem solving skills; therefore the software industry should pay more attention to this skill set. Most of the software parts are written by computer programmers independently without any assistance because larger piece of software is divided among smaller components. Therefore the ability to work independently with minimal support is a vital skill set for computer programmers, although there is the least demand for this skill set. Still there is a need to further acknowledge this skill set by the software industry. Changes are frequent in the software industry, and these changes range from specification and requirements, techniques and technology. Software programs are also subject to changes. Currently the software industry is not paying enough attention to the programmer's ability of being open and adaptable to changes. These skills set must be given more consideration, keeping in view the issues related to requirements management and the rapid growth in technology.

### 5.4 Software Tester

#### 5.4.1. Theoretical Foundation

Software testing involves identifying defects in a program and thus requires analytical and problem solving skills and insight. Software testers are considered as the worst enemies in the software development team because they are the group that gives the bad news of defects, an unenviable task which requires good communication and interpersonal skills to keep the team active and avoid conflicts [63]. Software testing is focused on finding defects, and there are many ways to make testing efforts more efficient and effective by using organizational skills. First, each module is tested on its own, isolated from the other components in the system. Such testing, known as unit testing, verifies that a module functions properly with the various input expected (and unexpected!) based on the module's design. After collections of modules have been unit-tested, the next step is to ensure that the interfaces among them are well defined. This is called integration testing. This requires many organizational skills to manage the testing process. System testing is the process of verifying whether all system components work properly together, a task which requires strong analytical skills. Testing strategies are neither random nor haphazard, but should be approached in a methodical and systematic manner. After a defect is detected, debugging can be a frustrating and emotionally challenging activity that may lead software engineers to restructure their thinking and revise their earlier decisions, this demands analytical and problem solving skills. Testing requires attention to detail, often performed by individuals working alone, and that requires the ability to work independently. Table 4 highlights the major requirements of the task that a software tester needs to perform during his job, and it also shows the possible mapping of soft skill to them. After mapping the soft skills required in carrying out these activities we found that communication, analytical and problem solving skills, interpersonal skills, the ability to work independently, and organizational skills are highly desirable for the job.

#### 5.4.2. Reflection from Software Industry

We collected the data of 165 software tester jobs currently advertised in the software job market, and analysis shows that only a single soft skill is in high demand: communication skills (78%). There is moderate demand for analytical and problem solving (42%), and organizational skills (39%). The least required soft skills for software tester are team player (27%), interpersonal skills (27%), the ability to work independently (15%), open and adaptable to changes (6%). Figure 8 summarizes the survey results. The analysis revealed that the software industry is not paying much attention to soft skills in the case of software testing. Only the ability to communicate effectively (communication skill) is in high demand, there is moderate demand for organizational skills, whereas some other important soft skills for a software tester such as analytical and problem solving, interpersonal skills, ability to work independently. Software tester performs manual execution of tests, recording of results, investigation and logging of faults. The job of the tester also requires collecting and documenting test requirements and, finally, producing test specifications. These activities are generally performed independently, therefore the tester should have the ability to work independently without or with only minimal supervision, the analysis shows that the software industry is least requiring of this skills set. The main challenge of the tester is to identify bugs and then convey this bad news to the people concerned. Sometimes the job of the tester become quite controversial because he is the one who lets the people know the bad news about the presence of a defect. This requires a great deal of interpersonal skills to handle conflicting situations. The software industry is not paying much attention to this skill set and currently this is one of the least required skill set. Therefore we suggest that the software industry should pay more attention to the skill sets of interpersonal skills and the ability to work independently.

**Table 4.** Software Testers' Job and Soft Skills Requirements

| Sample Software Tester Job Requirements | Possible Soft Skills Requirements |
|---|---|
| Gathers test requirements and produces test specifications. | Communication Skills, Analytical & Problem Solving Skills, Organizational Skills, Ability to work independently |
| Performs manual execution of tests, recording of results, investigation, and logging of faults. | Analytical & Problem Solving Skills, Organizational Skills, Interpersonal Skills, Ability to work independently |
| Demonstrates the ability to define and implement medium to large scale test plans and strategies according to quality objectives, project timelines and resources. | Communication Skills, Analytical & Problem Solving Skills, Organizational Skills |
| Manages defects, including the identification, logging, tracking, triaging, and verification of issues. | Analytical & Problem Solving Skills, Organizational Skills, Interpersonal Skills |
| Ensures test process, methodologies and tools are applied appropriately and that test phase entry/exit criteria are agreed upon by stakeholders and applied by the test team. | Communication Skills, Analytical & Problem Solving Skills, Organizational Skills, Interpersonal Skills |
| Maintains relevant test results databases. | Communication Skills, Analytical & Problem Solving Skills, Organizational Skills, Ability to work independently |
| Communicates and negotiates testing timelines, budget, staffing, scope and critical milestones with project managers. | Communication Skills, Analytical & Problem Solving Skills, Organizational Skills, Interpersonal Skills |

## 6. Discussion

The production of any piece of software involves a human element, requiring activities such as problem solving, analytical thinking, communication, and cognitive reasoning [64].Soft skills are usually overlooked in software engineering because the relationship between software engineering and soft skills is extremely complex and thus difficult to investigate. Nevertheless, it has been worthwhile studying which soft skills are required by the software industry and which are missing from the puzzle. Nowadays there are very few solo performers in most software organizations; people have to work collectively in teams of some sort, consequently, there should be a certain amount of diversity on the teams in terms of psychological type [65]. A broader conclusion of this survey in terms of the answer to the research question (RQ-II) put forward indicates that soft skills are in demand in the software industry, but only to some extent. Specifically only communication skills are given much attention by the software industry in all the four roles of software development whereas rest of the soft skills are still require enhanced understanding of their relationship with software development. This highlights a lack of understanding about the role soft skills play in the makeup of workers. Table 5 illustrates the overall picture of the survey's outcome and provides a solid image of the current status of the desirable soft skills in system analyst, computer programmer, software designer and software tester's roles.

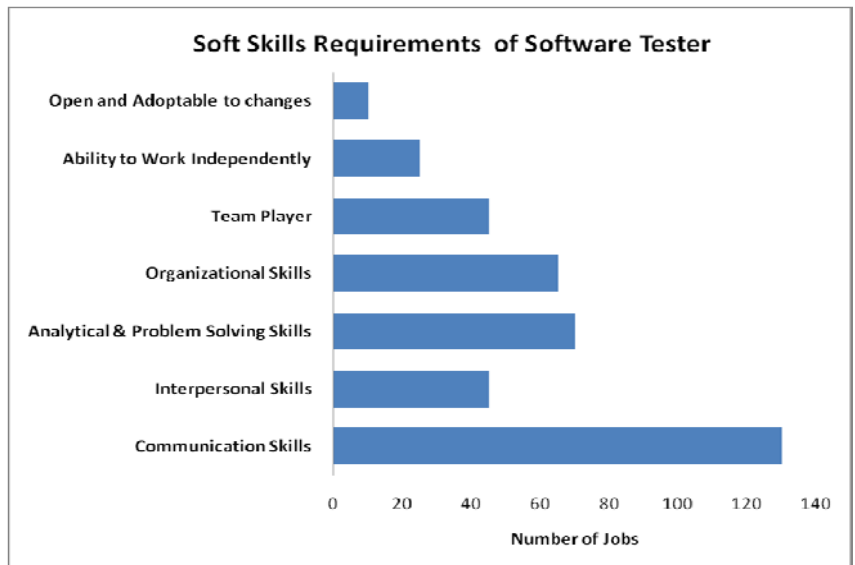

**Figure 8.** Soft Skills Desirable for a Software Tester

**Table 5.** Summary of the Survey Results

| Soft Skills | System Analyst | | | Software Designer | | | Computer Programmer | | | Software Tester | | |
|---|---|---|---|---|---|---|---|---|---|---|---|---|
| | H | M | L | H | M | L | H | M | L | H | M | L |
| **Communication skills** | √ | | | √ | | | √ | | | √ | | |
| **Interpersonal skills** | | | √ | | | √ | | √ | | | | √ |
| **Analytical and problem solving skills** | √ | | | √ | | | √ | | | | √ | |
| **Team player** | √ | | | √ | | | √ | | | | | √ |
| **Organizational skills** | | √ | | | √ | | | √ | | | √ | |
| **Ability to work independently** | | | √ | | | √ | | | √ | | | √ |
| **Open and adaptable to changes** | | | √ | | | √ | | | √ | | | √ |

( H: High Demand (> 66%), M: Moderate Demand (> 33% and ≤ 66), L: Low Demand (≤ 33%))

Communication skills have always been considered a vital set of skills in carrying out not only day to day operation in the job but also has been a significantly important tool when one has to liaise with customers. In software development, interaction with the stakeholders requires excellent communication skills from the employee. This interaction ranges from system analysts and end users to collect and manage requirements for the software. It can be prototyping by software designers and elaborating requirements to the users or programmers. Software testers need to communicate with the other roles of the software development process to plan, execute, and disseminate the results. The programmer translates the design into a computer program, a task which requires understanding of the requirements, design and further accelerates the requirements of good communication skills. The software industry values communication skills highly in all of the roles examined by the study. The ability to communicate effectively is in high demand in the job market.

Software development is a problem solving profession in which the customer highlights an issue in the current mode of operations and the software developer aims at providing a computer-oriented solution. Analytical and problem solving skills means that one has the ability to view and look into a situation from a very logical, systematic perspective and come up with a solution fitted to the scenario. In the process of software development each role, to a certain extent, has to make decisions which directly and indirectly have impact on the overall software cost, quality, and productivity. While he or she is making decisions some choices are challenging and take careful thought and consideration. In the survey results of this study, analytical and problem solving skills are in high demand for the role of systems analysts only. There is, however, a moderate demand for such skills in each of the other roles examined by the study. We would have expected to see that the demand for analytical and problem solving skills would have been high and common to all roles in software development, as the field is fundamentally a problem solving profession.

With the improving of software technology and the level of project management, the success of software project is not a special case. But this does not mean that software project can be finished easily [66]. IT projects have gained the universal attention from worldwide firms. However, because the skills of management related to exploitation and exploration are contradictory, firms still strive to find the effective methods to balance exploitative and exploratory activities [67]. Software is an area where things are changing rapidly as compared to other fields. For example, the programming methodologies have been evolved from first generation to fourth, and even now a fifth generation is coming into place. The development methodology has evolved from a single product development to a line of resulting products, giving rise to the popularity of a software product line. Similarly, in the case of users the changes are more frequent. One of the major reasons behind project delays and cost overrun in software projects is that requirements change. A soft skill such as being open and adaptable to changes is in low demand in all four roles examined. This is particularly disconcerting as the software field changes rapidly, as do the requirements and expectations of customers. Furthermore, advances in software design and innovative solutions can only be better developed by professionals who possess this subset of soft skills.

Organizational skills refer to the ability of individual managing tasks within time, resource and sequence of execution. This is significantly important skill in case of software development because one need to be very well plan in the execution of multiple tasks with in timeframe. Software development has always been criticized as over budget and over time project. Team members having well established organizational skills has the ability to reduce over budget and time. This survey reports a moderate demand of organizational skills in case of system analysts, programmer and software tester whereas a low demand in case of software designer. The job of software designer needs further acknowledgement to this vital skill as fundamentally designer is involved in breaking apart the whole software into smaller manageable pieces. The designer has to ccollaborate closely with management, engineers and fellow designers to evaluate and iterate on ideas and designs. These iterations and top down approach requires considerable amount of organizational skills.

Interpersonal skills refer to how people relate effectively to one another in pursuit of common goals. Software development is not a one-man show, it requires team efforts. When people are interacting in teams they have different opinions and styles. This raises the possibility of conflicts among team members which can later affect the productivity of the team. Positive interpersonal skills of the team members increase the productivity of the organization, since the number of conflicts is reduced and trust increased. Some roles of the software development phase also require interaction outside the boundary of the development team. For instance, the key role of system analysts requires interacting with customers who may not be proficient enough in computing to gather and elaborate the requirements of the project. The job of system analyst requires a good set of interpersonal skills to handle the situations. The findings of this survey illustrates that none of the job role in software development requires this skill in high demand. Only computer programmer's job advertisements show a moderate demand for interpersonal skills, while the role of systems analyst, software designer and software tester shows little requirement for this skill in the survey reported here. We found this surprising, as both systems analysts and software testers are required to interact with a broad range of individuals and teams as part of their basic role.

Although, as we mentioned earlier, software development is a team effort, but if we look carefully further into some of the roles, they require an individual's having the ability to work independently as well. The ability to work independently requires self-awareness, self-monitoring, and self-correcting skills. It shows that one takes the initiative rather than waiting to be told what to do. It further highlights one's ability to do what is asked to the best of one's ability, without the need for external prodding, and also one's ability to work until the job is completed. For example, in the case of a computer programmer, one has to write code independently and with minimal support to cover the functionality. Similarly the role of software tester also requires the ability to work independently quite often because one has to write, execute, and report test cases to do various types of testing. The ability to work independently is in low demand for system analyst, designer, programmer, and tester's roles, as reported here in this survey. This finding is surprising, as the roles of a software tester and programmer are generally solitary whilst designers and analysts tend to involve more team-based activities.

## 7. Conclusion

This survey reports some of the vital statistics from industry about the requirements of soft skills in various roles of software development phases. The result of this investigation reveals that currently the software industry is paying attention to soft skills while hiring, but there is a need to further acknowledge the role of these skills in software development. We conclude that like any other job the software development takes a variety of soft and hard skills to solve the myriad problems related to software development and industry should understand this complex relationship. It might be suggested that organizations would be well-served by a conscious attempt to diversify the soft skills of their software engineers. Therefore all soft skills are important to software development and can make a contribution to tackling the so-called software crisis. Further exposure to and enhanced understanding of soft skills from the software industry can help an appreciation of this fact to flourish. Achieving such a variety of skills would enable us to bring a richness of talent and points of view to bear upon the inherent complexity of software systems.